\documentclass[aps,prb,preprint,groupedaddress,showpacs,footinbib]{revtex4-2}
\pdfoutput=1
\usepackage{bm}
\usepackage{graphicx}
\usepackage{mathrsfs}
\usepackage{amssymb}
\usepackage{amsmath}
\usepackage{booktabs}
\usepackage{diagbox}
\newcommand{\de}{\mathrm{d}}

\begin{document}
\title{Origins of anomalous low Raman exponents in single-molecule magnets}

\author{Lei Gu}
\affiliation{Department of Physics and Astronomy, University of California, Irvine, California 92697, USA}
\author{Ruqian Wu}
\email[]{wur@uci.edu}
\affiliation{Department of Physics and Astronomy, University of California, Irvine, California 92697, USA}

\begin{abstract}
The Raman exponent of single-molecular magnetic relaxation may take various unexpected values because of rich phonon spectrum and spin-phonon coupling of molecular crystals. We systematically examine the origins of different abnormalities, and clarify misunderstandings in the past, particularly the appropriateness of the fitting procedures for the exponents. We find that exponential laws raised by optical phonons can yield spurious power laws with low exponents. This observation indicates long-standing misunderstandings for origins of low Raman exponents in a large bulk of single-molecule magnets. Resulting from spin-lattice coupling with optical modes, presence of these exponents suggests the importance of the local dynamical environment for the magnetic relaxation in this regime.
\end{abstract}
\maketitle

\section{Introduction}
The spin-vibration and spin-phonon couplings in solids and molecules are widely discussed but are puzzling in many cases such as magnetic phase transition, damping and relaxation. These issues are even more significant and complex in dealing with single-molecule magnets (SMMs) which have received increasing attention as possible qubits~\cite{Zadrozny2015,Atzori2016,Hu2018} for quantum information processing and storage~\cite{Graham2017,Escalera2018,Gaita2019,Atzori2019}. In fact, the control of spin-vibration coupling in SMMs by selecting appropriate ligands and substrates is the most viable strategy to lower their relaxation rate or equivalently to extend their quantum coherence time. Obviously, it is necessary to clarity the effect of vibrational excitations on spin relaxation of SMMs such as via the Raman process, so that comprehensive understanding in their quantum behaviors and practical design rules for SMM-based devises can be established.

In the past three decades, considerable advancements have been made for the synthesis and characterization of complex molecules and molecular solids~\cite{Ardavan2007,Magnani2010,Harman2010,Freedman2010,Zadrozny2011,Lucaccini2014,Gomez2014,Bader2014,Atzori2016b,Moseley2018,Rajnak2019,Zadrozny2013,Blagg2013,Chen2016,Goodwin2017,Guo2018}.
The judiciously designed dysprosoceniums~\cite{McClain2018,Guo2018,Goodwin2017,Ding2016} show magnetic hysteresis at the liquid nitrogen temperature, indicating the possibility for the use in SMM-based devices. Single magnetic molecules were used to functionalize tips of scanning tunneling microscopy (STM) for measuring and mapping exchange interactions with a sub-Angstrom spatial resolution~\cite{Bartels1997,Lee1999}. Strong intermixing between vibrational and spin excitation were also directly detected in the inelastic electron tunneling spectrum (IETS)~\cite{Czap2019}, which provides a useful tool to quantitatively investigate and engineer molecular magnetic systems.

However, on the theoretical side some long-standing puzzles are still not well understood and call for fundamental studies. One of the outstanding issues is the presence of anomalous Raman exponents (see e.g.~\cite{Harman2010,Zadrozny2013b,Guo2018,Ding2018,Rajnak2019,
Wang2019,Kobayashi2019,Cui2019,Vallejo2019,Handzlik2020}) that generally deviates from the standard values~\cite{Abragam2012,Shrivastava1983}. Considering that the magnetic hysteresis usually occurs jointly with the dominance of Raman relaxation~\cite{Guo2018,Goodwin2017}, the Raman regime appears to be suitable for computing and sensing applications and deserves careful investigations. Our systematic examine suggests that the abnormality of these low Raman exponents is due to mistakenly taking exponential laws as power laws. As the exponential laws arise from coupling with local vibrational modes, the finding highlights the important role of local vibrational modes for the magnetic relaxation of SMMs, and general guidances for lengthening the relaxation time can be drawn accordingly.

\section{Inappropriateness of the conventional Raman exponents}

\begin{figure}
\includegraphics[width=0.6\textwidth]{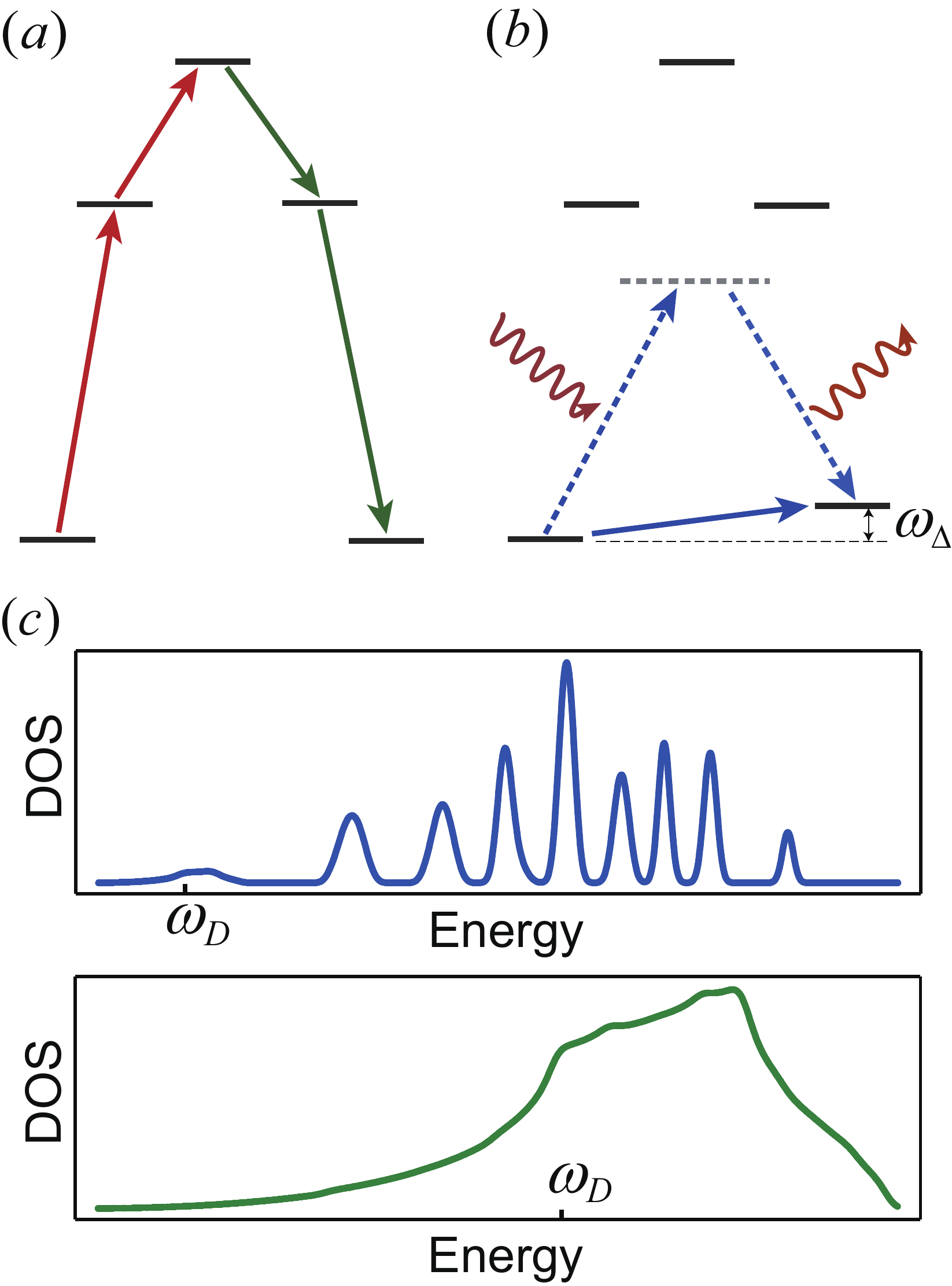}
\caption{(a) When the relaxation process is mediated by excitation states, Raman processes of these transitions do not yield power law dependencies, which can only result from direct tunneling between the ground state doublet as shown in (b). (c) Phonons of a SMMs system (upper) are made of acoustic phonons with very low energy and optical phonons from broadening of local vibrational modes; this difference with bulk materials (lower) brings about several peculiarities to the Raman processes in SMMs, and gives rise to the abnormalities.}\label{tunnelling}
\end{figure}

Most SMMs designed for slow magnetic relaxation have strong uniaxial magnetic anisotropy (see Ref.~\cite{Gomez2014} for an exception), described by $H_S= -D S_z^2-E(S_x^2-S_y^2)$ with $D\gg E$. This sets an effective barrier $U_{eff}=DS^2$ for the standard Orbach relaxation pathway as sketched in Fig.~\ref{tunnelling}(a). In Ref.~\cite{Gu2020}, we clarified that the Raman processes for the transitions in this pathway cannot lead to power laws~\cite{Goodwin2017}, as the Orbach barrier set the time scale $\tau = \tau_0e^{U_{eff}}$ and the Raman processes somewhat modify the prefactor $\tau_0$. This implies that the power laws can only arise from the direct tunneling between the ground state doublet. The significant Raman process conventionally referred to should be the one shown in Fig.~\ref{tunnelling}(b), that is, direct tunneling mediated by simultaneously absorbing and emitting of a phonon.

At high temperature, all spin states are well accessible, so the Orbach process dominates the magnetic relaxation. When the temperature is reduced, the Raman process can be dominant and power laws emerge. The intriguing cooccurrence of magnetic hysteresis and emergence of the power law can be understood by the change of the $\tau-T$ dependence itself. In the Orbach regime, a small temperature reduction can dramatically increase the relaxation time, especially for large Orbach barriers. When reaching the Raman regime, this sensitivity is significantly weakened because of the transition to a power law $\tau-T$ relation (cf. Eq.~(\ref{time})). When experimentalists try to make a tradeoff between high temperature and long relaxation time, this transition point is likely to be “selected” as the emergence point of magnetic hysteresis. Since only the ground state doublet is involved in the spin dynamics, another merit of the Raman regime for practical applications is the purity, i.e., the system is a desired two state qubit.

With relaxation times ($\tau$) of a SMM in a wide temperature range, typical $\tau-T$ curves can be fitted by a relation
\begin{equation}
\tau = \tau_{0}e^{\frac{U_{eff}}{k_BT}}+ CT^{-n}.
\label{time}
\end{equation}
The first term represents the Orbach process, and the second term is mainly due to the Raman process. The standard Raman exponent at low temperature should be $n=7,8,9$~\cite{Abragam2012,Shrivastava1983}. Nevertheless, a large bulk of observations gives $n= 3\sim5$ (see e.g.~\cite{Harman2010,Zadrozny2013b,Guo2018,Ding2018,Rajnak2019,
Wang2019,Kobayashi2019,Cui2019,Vallejo2019,Handzlik2020}). These unconventional values are usually left unexplained or ascribed to the optical-acoustic mechanism~\cite{Singh1979}. However, most of SMMs with slow magnetic relaxation was designed to have strong uniaxial magnetic anisotropy ($D \ll E$), so that their Orbach barriers are high. This makes the splitting between the ground state doublet small and implies that the absorbed and emitted phonons in the Raman process should be of the same type. As a result, the optical-acoustic mechanism is inapplicable, and other mechanisms should be explored for the explanation of anomalous Raman exponents for the spin relaxation in SMMs.

We found that the profile of phonon density of states (DOS) of SMM systems may lie at the heart of these abnormalities. Since the magnetic measurements are usually performed on molecule crystals consisting of the magnetic complexes and solvent molecules, we used a 3D oscillator model to show generic traits of the phonon DOS. Using a $3\times3\times3$ supercell, and assuming that the intra-cell ionic force constant is one order bigger than the inter-cell van der waals type ones, the phonon DOS of typical molecular crystals is generated as given in the upper panel of Fig.~{\ref{tunnelling}}(c). The comparison between the phonon DOS curves of ordinary crystals (lower panel) clearly demonstrates the reason why the conventional Raman exponents cannot arise in SMMs. The derivation relies on extending the integration limit $\omega_D/T$ of the Debye integral (cf. Eq.~(\ref{rate})) to the infinity. However, because $\omega_D$ is small in typical SMMs systems, such extension is applicable only at very low temperature. Here, we note that due to closeness of the energy levels, the peaks are broadened. In concrete material, the optical phonon peaks can be much narrower as shown in Ref.~\cite{Chiesa2020}. In the following, we will investigate, from three aspects, how the small Debye energy and the discrete nature of optical phonons affect the Raman exponent.

\section{Origins of anomalous Raman exponents}
From the second order spin-phonon coupling Hamiltonian, the tunneling rate can be derived as
\begin{align}
p =  N(\omega_{\Delta})\iint \frac{\pi|a_{qq'}|^2}{2\omega_{q}\omega_{q'}}\de\omega_{q}\de\omega_{q'}\rho(\omega_{q})\rho(\omega_{q'}) \{&[N(\omega_{q})+N(\omega_{q'})+1]\delta(\omega-\omega_{q}-\omega_{q'})\nonumber\\
+ &[N(\omega_{q})-N(\omega_{q'})]\delta(\omega+\omega_{q}-\omega_{q'})\},
\label{rate}
\end{align}
where $\omega_{q}$ denotes phonon frequency, $N(\omega_{q})$ the Bose-Einstein distribution, $\rho(\omega_q)$ the phonon DOS, and $\omega_{\Delta}$ the energy difference between the ground state doublet as shown in Fig.~\ref{tunnelling}(b). By energy conservation, we may identify the first term as the double phonon process whereby two phonons are absorbed, and the second terms as the Raman process whereby a phonon is absorbed ($\omega_{q'}$) and a phonon of lower energy is emitted ($\omega_{q}$). Here, we include the double phonon process since it is naturally derived from the Hamiltonian of second order spin-phonon coupling. In other words, the Raman process is inevitably accompanied by the double phonon process. We showcase the requirement of conventional exponents with the non-Kramers case. One may refer to Refs.~\cite{Shrivastava1983,Gomez2014} for the derivation and discussion on the magnetic relaxation of Kramers systems.

In the long wavelength limit, the continuum mechanics applies~\cite{Blume1962,Abragam2012}, which implies that the lattice deformation caused by acoustic phonons is approximately proportional to the phonon momentum. As phonon-spin coupling essentially reflects variation of electronic state due to the lattice deformation, this proportionality applies to the coupling strength, i.e., $|a_{qq'}|\propto |q||q'|$. According to the Debye dispersion $\omega_q\propto |q|$, the coupling coefficient can be approximated as $|a_{qq'}|\propto\omega_q\omega_{q'}$. Together with Debye phonon DOS $\rho\propto \omega^2$, the second term of Eq.~(\ref{rate}) gives the Debye integrals for the Raman process. The standard Raman exponents arise when the integration limit $\omega_D/k_BT$ is extended to the infinite. As a result, we obtain $p\propto T^6 N(\omega_{\Delta})$. High temperature or small $\omega_{\Delta}$ expansion of $N(\omega_{\Delta})$ results in $\tau^{-1} \propto p\propto T^7/\omega_{\Delta}$, the standard relation for the non-Kramers doublets.

The requirement $\omega_D\gg k_BT$ for legitimacy of the integration limit extension, however, is limited to very low temperature in SMMs. A small Debye energy relative to the temperature is the first aspect that gives rise to anomalous Raman exponents. Because of the weak inter-molecular interaction, the acoustic phonons mainly represent the inter-molecule motion~\cite{Gu2020}. Assuming that the inter-molecular interaction is one order weaker than the intra-molecular interactions and masses of the molecules are one order larger than an ordinary atom, the Debye energies of SMMs are one order smaller than those of ordinary crystals, and $\omega_D \approx 20$ cm$^{-1}$ is a representative estimation. As $\omega_D=20$ cm$^{-1}$ amounts to $28.7$ K, the conventional Raman exponent is appropriate when the temperature is well below $10$ K.

\begin{figure}
\includegraphics[width=0.95\textwidth]{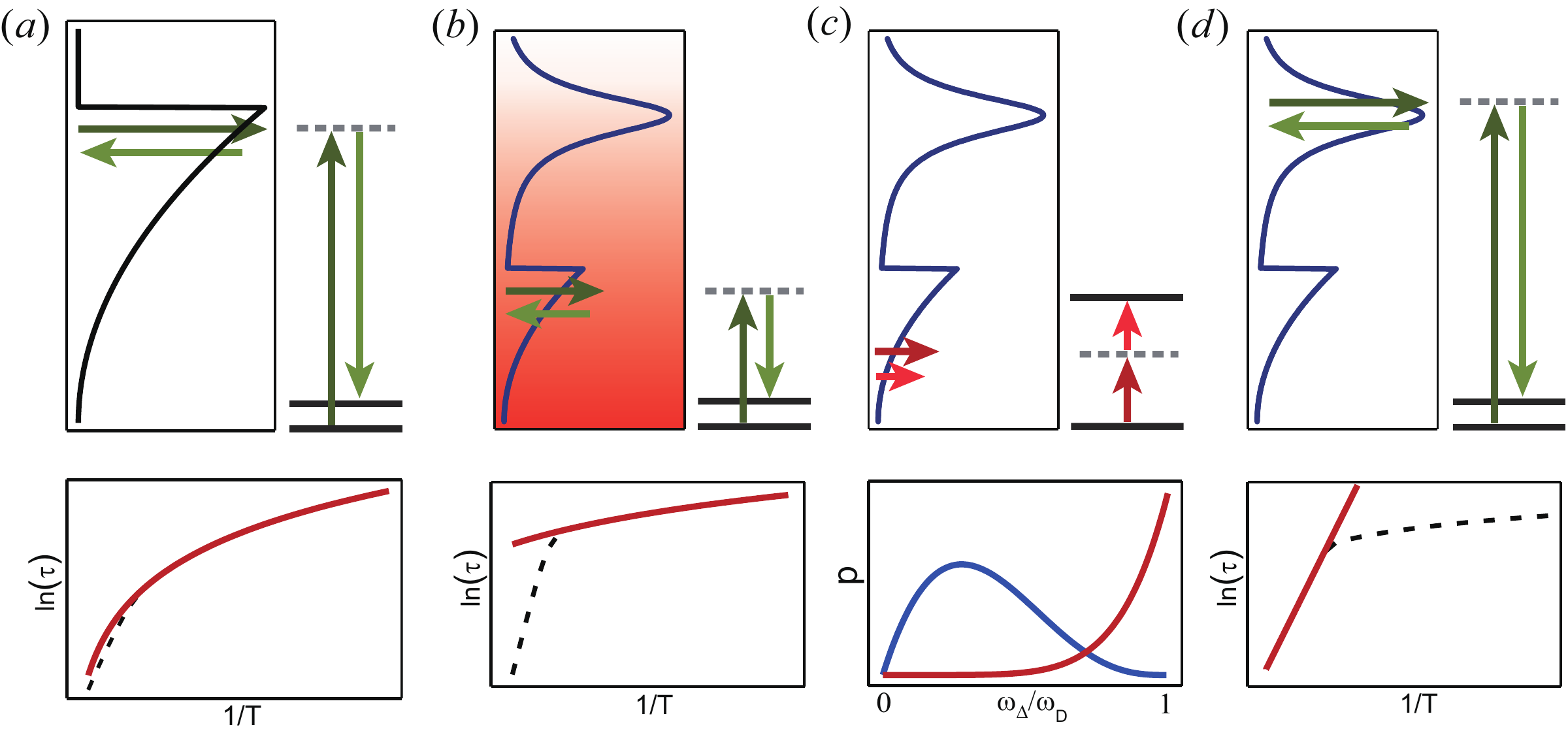}
\caption{(a) The conventional Raman process is mediated by acoustic phonons in solids having a large Debye energy. (b) Because of a small Debye energy, the acoustic phonons in SMM systems are well accessible for relatively high temperature, leading to $\tau^{-1}\propto T^2$. (c) For SMMs with large angular momentum, the Zeeman splitting can be comparable with the Debye energy, and the double phonon process (red) can perceivably contribute to the tunneling rate, even surpassing the Raman process (blue). (d) Raman process due to a local vibration yields an exponential dependence, which is a major cause of anomalous Raman exponents.}\label{mechanisms}
\end{figure}

On the contrary, in SMMs the condition $\omega_D \ll k_TB$ can be well satisfied, which implies that the Debye phonons are well accessible (Fig.~\ref{mechanisms}(b)). With high temperature expansion of the Bose-Einstein function $N(\omega_q)$, the second term of Eq.~(\ref{rate}) gives $\tau^{-1}\propto T^2$. As shown in~\cite{Chiesa2020}, this $T^2$ dependence is general and also applies to Kramers systems. The observation $n=2.15$ in Ref.~\cite{Goodwin2017} is a clear case of high temperature Raman process. Phonon DOS not in perfect Debye form and variation of spin-phonon coupling strength with momentum may cause small deviations. In bulk materials, due to large Debye energies, the high temperature expansion is rarely used. But in SMMs, the high temperature Raman process can be essential and constitutes an origin of small Raman exponents.


In the double phonon process, transitions among spin states are accompanied by emission or absorption of two phonons. Because of the energy conservation $\omega_q+\omega_{q'}=\omega_{\Delta}$, only phonons in the range $[0, \omega_{\Delta}]$ can contribute to the direct tunneling. In contrast, phonons in the range $[0, \omega_D]$ participate in the Raman relaxation process. For small splitting and large Debye energy ($\omega_{\Delta} \ll \omega_D$), the double phonon process is inconsiderable compared to the Raman process and has rarely been mentioned. As $\omega_D$ is small in SMMs and the Zeeman splitting may be sizable for large spins, the relative contribution of these two processes are worth of careful investigation.

Our numerical estimation shows that for $\omega_D = 20$ cm$^{-1}$, the double phonon process surpass the Raman process when $\omega_{\Delta}/\omega_D \gtrsim 0.7$ (Fig.~\ref{mechanisms}(c)). At very low temperature ($T\lesssim 4$ K), only the low energy phonons are effective for both processes and the phonons in the range $[\omega_{\Delta}, \omega_D]$ are less important, so the critical ratio can be largely reduced (see supplementary). In general, because of the small splitting between the ground state doublet, the double phonon process can be safely neglected for SMMs with strong uniaxial magnetic anisotropies. For example, a $1000$ Oe magnetic field yields an $1.4$ cm$^{-1}$ Zeeman splitting between $|\pm S\rangle$ for $S=15/2$. As $\omega_{\Delta}/\omega_D \ll 1$, this does not lead to strong double phonon processes. Therefore, the double phonon process is still an insignificant relaxation channel, except for extremely strong magnetic field and low temperature, which results in large $\omega_{\Delta}$ or small critical $\omega_{\Delta}/\omega_D$ ratio, respectively.


Most recently, the mechanism of under-barrier relaxation in absorbate magnetic atoms~\cite{Donati2020} and SMMs~\cite{Harman2010,Freedman2010,Vallejo2012,Coca2013,Zhu2013,Fataftah2014,
Pedersen2015,Novikov2015,Rechkemmer2016,Rajnak2019,Wang2019,Kobayashi2019,Krylov2017,Flores2019} has been explained~\cite{Donati2020,Gu2020}. It is found that the second order Raman process due to a vibrational mode can yield exponential temperature dependence $\tau = \tau_0 e^{U_{vib}/ k_BT}$ as shown in Fig.~{\ref{mechanisms}(d)}. This means that the vibrational mode raises an effective relaxation barrier equal to its energy. In the appendix we show that vibronic barriers can also result from the Raman process due to the first order spin-phonon coupling to the second order perturbation. Since the Raman transition between a doublet due to the first order spin-phonon coupling is mediated by another spin state, it is not limited by the time reversal symmetry. Therefore, the vibronic barrier can exit in both non-Kramers and Krammers SMMs. The exponential form implies that the conventional use of Eq.(\ref{time}) for fitting is a long standing misstep, since the actual temperature dependence is an exponential function or summation of a series of them. When forcefully fitting it with a power law $\tau \propto T^{-n}$, exponents unrelated to the conventional Raman process might be obtained. This vibronic barrier is the third aspect concerning the anomalous Raman exponents, supposedly the most significant one.


\begin{figure}
\includegraphics[width=0.95\textwidth]{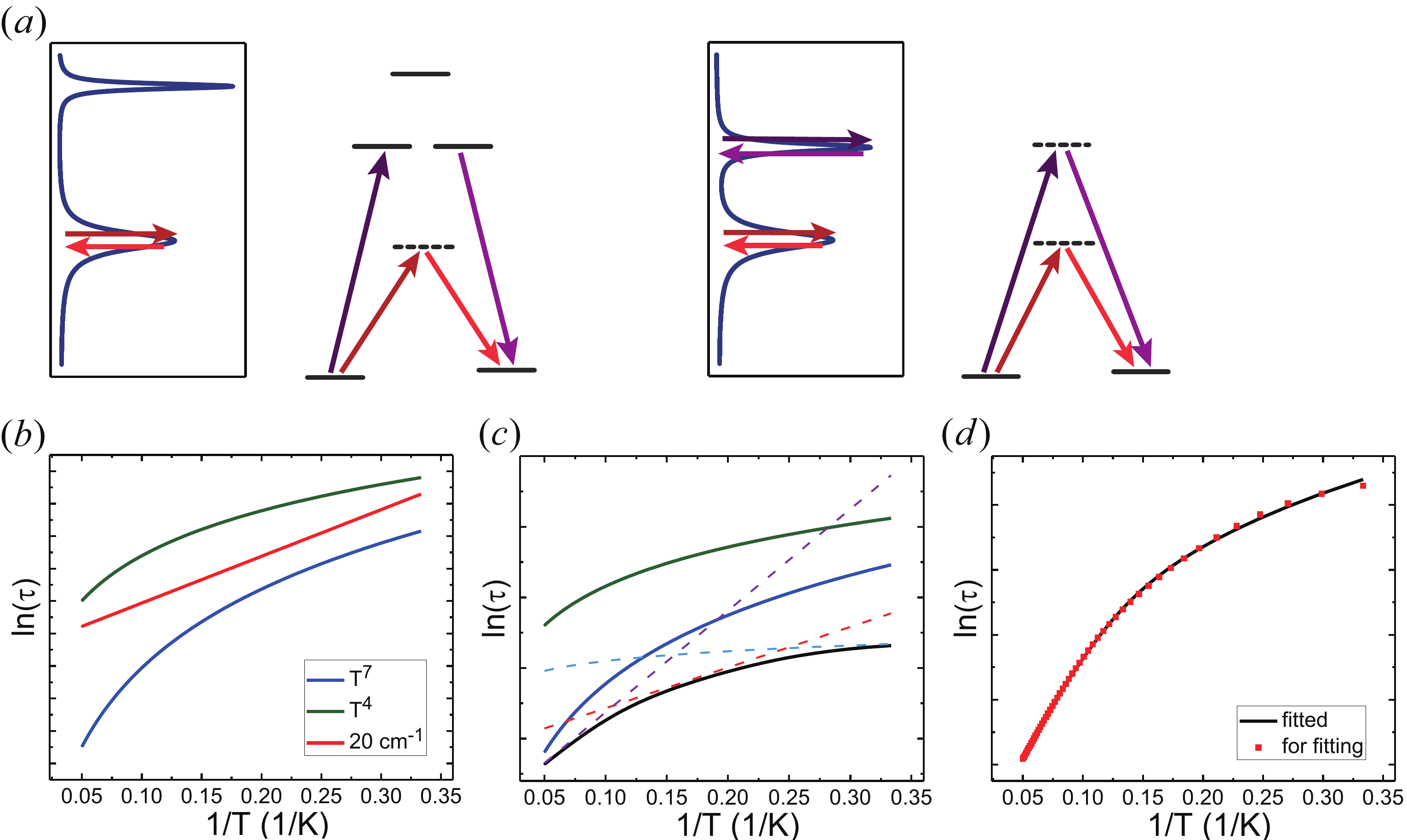}
\caption{(a) Two cases where small Raman exponents could be deduced from Eq.(\ref{time}). (b) Exponential dependence of an energy barrier $20$ cm$^{-1}$ is similar with $\tau^{-1}\propto T^4$, and more clearly different with $\tau^{-1}\propto T^7$. (c) When summed up with exponential dependence of a higher energy barrier and $\tau^{-1}\propto T$ for the direct process, the curve (black) can be quite deceptive and easily mistaken as $\tau\propto T^4$. (d) In this example, such a summed curve can be well fitted by Eq.~(\ref{time}) with exponent $n=4.02$.}\label{fitting}
\end{figure}

We take a vibrational mode $\omega=20$ cm$^{-1}$ for instance. As shown in Fig.~\ref{fitting}(b), the curve of $\tau^{-1}\propto e^{20/K_BT}$ has similar variation range with $\tau^{-1}\propto T^4$ than with $\tau^{-1}\propto T^7$, the standard value for non-Kramers doublet. Note that the Raman process mediated by his mode is not the only relaxation channel. At high temperature, the relaxation is dominated by the Orbach process (left Fig.~\ref{fitting}(a)) or another intra-molecular vibrational mode (right Fig.~\ref{fitting}(a)) with stronger coupling with the spin than this mode. This adds another exponential function as denoted by the doted purple line in Fig.~\ref{fitting}(c) (here a $50$ cm$^{-1}$ barrier is assumed). In the other side, the direct process can be dominant at low temperature and raises power law $\tau\propto T^{-1}$~\cite{Abragam2012,Shrivastava1983}. When these two additional functions are included, we have the black curve in Fig.~\ref{fitting}(c), which is a generic curve from most experimental measurements. Compared with $\tau\propto T^{-4}$, it is quite deceptive and can be easily mistaken (here the position is not important due to the ln$(\tau)$ form). In Fig.~\ref{fitting}(d), we fit the curve with Eq.~(\ref{time}), which leads to an exponent $n=4.02$. Typically, there are many vibrational modes of SMMs in the range $10\sim 30$ cm$^{-1}$. The variations of the mode energies and relative contributions from the three relaxation processes are expected to cause derivations from $n=4$. Therefore, the improper fitting with Eq.~(\ref{time}) may give diverse exponents in the range $3\sim 5$.

\begin{figure}
\includegraphics[width=0.5\textwidth]{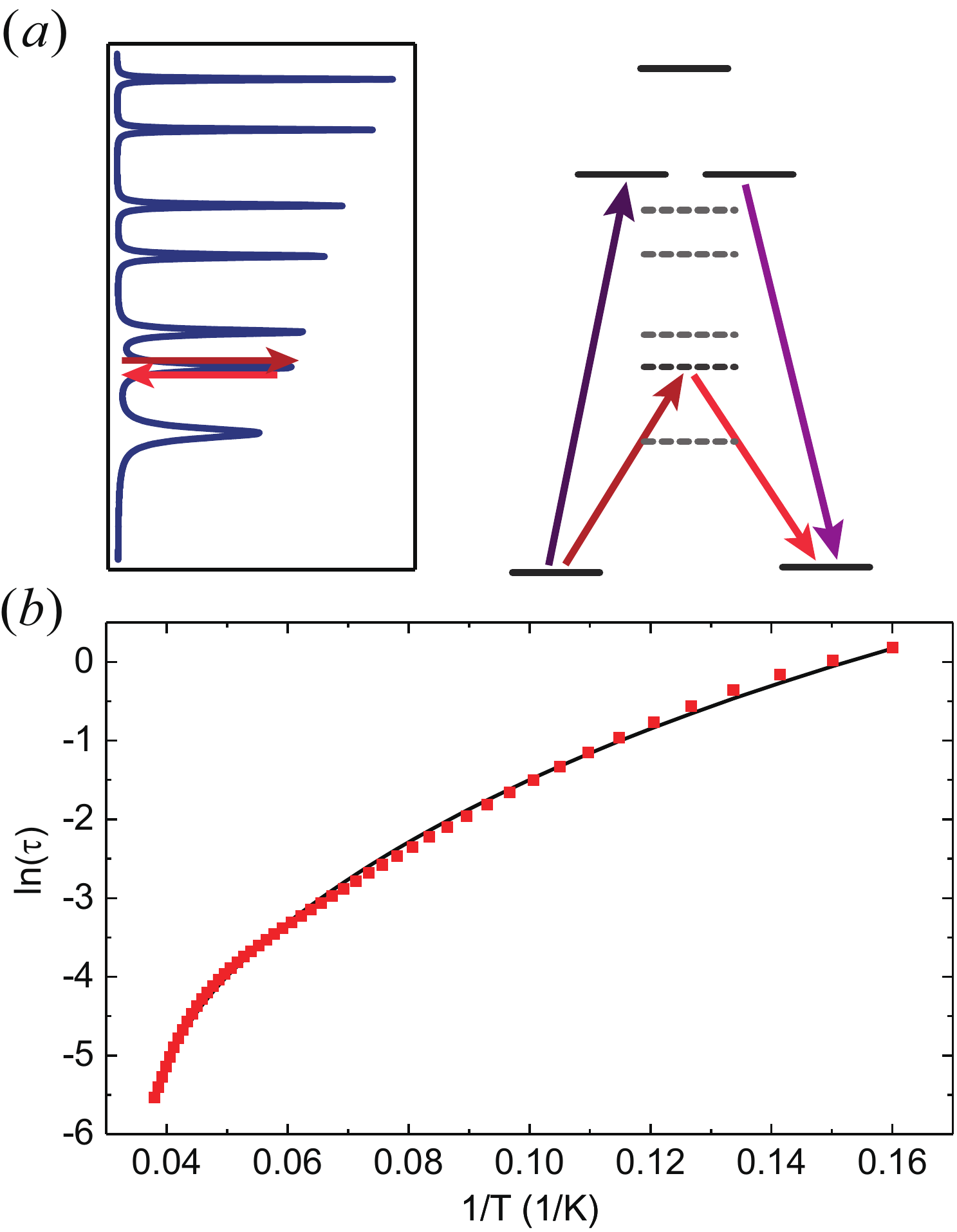}
\caption{(a) When the Orbach barrier is high, the vibrational modes below it collectively contribute to the Raman relaxation. (b) Summation of a series of exponential functions still leads to deceptive curves that may be mistaken as a power function. Using Eq.~(\ref{time}), the data (red) can can be fitted with an exponent $n=3.55$.}\label{real1}
\end{figure}

The above case applies when the zero field splitting is small and only a few vibrational modes have lower energies. When the zero splitting is large, many vibrational modes having energies lower than the Orbach barrier may collectively contribute to the Raman relaxation (Fig.~{\ref{real1}}(a)). To validate our approach, we calculated the vibrational modes of the Co-NCCN metallacycles with the ORCA package~\cite{Neese2018}. According to Ref.~\cite{Rechkemmer2016}, it is a spin-3/2 molecule with uniaxial anisotropy $D=115$ cm$^{-1}$, i.e., an Orbach barrier of $230$ cm$^{-1}$. Summing up the exponential function for $24$ modes below the Orbach barrier, we obtained the curve in Fig.~\ref{real1}(b), which is close to the experimental result. Fitting the curve with Eq.~(\ref{time}) gives a Raman exponent of $n=3.55$. The first order and second order spin-phonon coupling (partial derivative of $D, E$ w.r.t. atomic displacement) are assumed to be in the order $0.1$ cm$^{-1}/$\AA\  and $0.01$ cm$^{-1}/$\AA$^2$, consistent with the typical values~\cite{Lunghi2017} (see supplementary for details for the parameter estimation).

\begin{figure}
\includegraphics[width=0.6\textwidth]{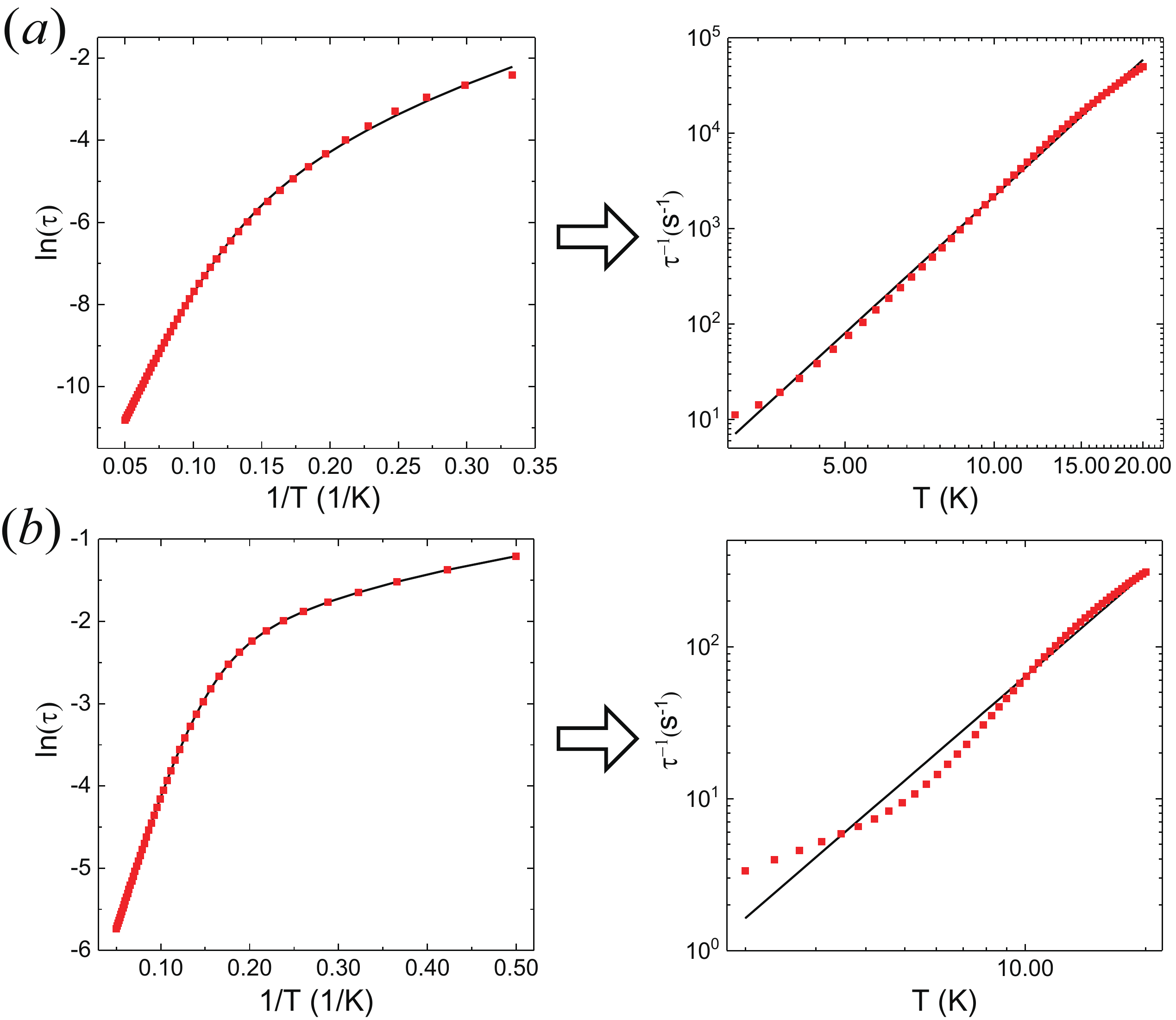}
\caption{(a) Fitting the data in Fig.~\ref{fitting}(d) with $\tau\propto T^{-n}$ and log-log scaling gives $n=7.7$; The closeness to the standard Raman exponents may cause misinterpretation of the underlying mechanism. (b) For lower vibronic barriers, small exponents can be obtained; for instance, a barrier of $25$ cm$^{-1}$ leads to an exponent $n=2.3$.}\label{real2}
\end{figure}

Unaware of the mechanism of vibronic barrier, sometimes the experimental data were forcefully fitted by a power law with log-log scaling and unreasonable values of $n$ might be obtained. For example, if we fit the data in Fig.~\ref{fitting}(d) with logarithm scaling for both $\tau$ and $T$, we have an exponent $n=7.7$, as shown in Fig.~\ref{real2}(a). When the dominant vibrational mode has lower energies, the exponent can take smaller values. For instance, in Fig.~\ref{real2}(b), the Raman process of an vibronic barrier of $25$ cm$^{-1}$ together with the direct process leads to an exponent $n=2.3$ by the log-log fitting. Since this fitting procedure can lead to exponents close to the conventional values, special care should be taken when using it. As argued in the proceeding, the conventional Raman process takes effect at very low temperature. If the linearity extends over $10$ K, it is more likely that the vibronic barriers are in play.

\section{Discussions and conclusion}

We have investigated mechanisms that can lead to unusual Raman exponents, and fitting procedures that may mistake them. Except the high temperature exponent $n\approx 2$, one should take special care when using the power law to fit the Raman process. Ab initio calculations and suitable experimental procedures are needed to reveal the dominate relaxation channel. Based on generic properties of SMMs, some general trends can be inferred. As seen in Fig.~\ref{tunnelling}(c), the acoustic phonons have small DOS. Moreover, because the intra-molecule deformation is weak for acoustic phonons, the spin-phonon coupling is usual much weaker than the optical phonon. Although the optical phonons have higher energies, it is quite likely that they dominate over the acoustic phonons at relatively high temperature. Since both the first and second order Raman process present for non-Krammers doublet, the situation becomes even more complicated. Measurements such as the field dependence of the relaxation time can help distinguishing the two type of Raman processes.

For the heavy lanthanide metallocenium cations~\cite{Goodwin2017b} that recently receive much attention, there is an opinion that the pseudo spin Hamiltonian like $H_S= -D S_z^2-E(S_x^2-S_y^2)$ is no longer applicable due to the strong spin orbital coupling. However, as the spin operators satisfy the commutation relation of angular momentum and their products form a basis (i.e., the Stevens operators~\cite{Stevens1952}), the electronic states in these cations can still be described in the form of the pseudo spin Hamiltonian. For the irregular eigen energies that significantly deviate from the hyperbolic form ($-DS_z^2$), the reason is not inapplicability of the pseudo spin formulation but large $E$ and nonnegligible high order magnetic anisotropies. This perspective provides a clear and unified explanation to why most of the cations except the dysprosoceniums in Refs.~\cite{Goodwin2017,Ding2016} fail to achieve long relaxation time and molecular magnetic hysteresis. On one hand, these terms mix states $|S_z \rangle\ (|S_z| \ll S)$ into $|\pm S\rangle$ with sizable portions to form the ground state doublet, or even result in ground state doublet no longer based on $|\pm S \rangle$. On the other hand, coupling between the high order magnetic anisotropies and phonons makes transitions between states with $\Delta S_z > 2$ possible. These two factors can lead to shortcut of relaxation pathways via transitions among the low energy spin states, which significantly shorten the relaxation time and make high temperature magnetic hysteresis unachievable. Noting that the calculated tunneling rates for the dysprosoceniums and the other cations do not show a clear magnitude difference~\cite{Goodwin2017b}, this argument may indicate inaccuracy of the ab initio calculations for atoms with strongly localized electronic states, calling for development of numerical techniques.

In summary, because the existing theories can give rise to various Raman exponents, the origins of anomalous low Raman exponents have eluded researchers’ attention in the study of SMMs. This leads to misuse of the power laws for fitting the $\tau-T$ dependence in many SMMs. Because of complexity in the Raman regime, the fitted result can only provide ambiguous information about the underlying relaxation mechanisms, for which more detailed and further measurements are needed. In general, as the optical phonons tend to dominate over the acoustic phonons at relatively high temperature, it is probably the dominant relaxation channel near the transition point from the Orbach to the Raman regime, which may be best suitable for practical application. Engineering of the local dynamical environment~\cite{Atzori2016,Moreno2017,Escalera2018,Parker2020,Yu2020,lunghi2020} should be one of our major concerns, so that the spin-lattice coupling can be weakened and low energy vibrational modes can be avoid.


\section*{acknowledgements}
The work is supported by the Department of Energy (grant No. DE-SC0019448) and computing time by NERSC.

\appendix
\section{Vibronic barriers due to the first order Raman Process}
We start from the Eq.~(50) of Ref.~[17]. The transition rate from State $|a\rangle$ to the other state $|b\rangle$ in the doublet via an intermediate state $|c\rangle$ is given by
\begin{equation}
P_{ba}=\frac{2\pi}{\hbar}\Bigg| \frac{\langle b, n_k-1, n_{k'}+1|H_1|c, n_k-1, n_{k'}\rangle \langle c, n_k-1, n_{k'}|H_1|a,n_k, n_{k'} \rangle}{E_c-E_a-\hbar\omega_k}\Bigg|^2\delta(E_b-E_a-\hbar\omega_k+\hbar\omega_{k'}).
\end{equation}
We consider the ground state doublet of strong axial SMMs, $E_b-E_a\ll 1$ cm$^{-1}$, which requires that $\omega_{k}, \omega_{k'}$ is around the same peak for a local vibrational mode. The spin-phonon coupling is given by
\begin{equation}
H_1 = \frac{\partial H_{spin}}{\partial V_{\alpha}} \sqrt{\frac{\hbar}{2\omega_{\alpha}}}(b_{\alpha} + b^{\dagger}_{\alpha}),
\end{equation}
where $V_{\alpha}$ denote the displacement for mode $\alpha$. Integration over DOS peak of this mode gives
\begin{equation}
P_{ba} = C \iint \frac{2\Gamma_{\alpha}}{(\omega_{k'}^2-\omega_{\alpha}^2)^2+\Gamma_{\alpha}^2} \frac{2\Gamma_{\alpha}}{(\omega_k^2-\omega_{\alpha}^2)^2+\Gamma_{\alpha}^2}\frac{1}{(E_c-E_a-\hbar\omega_k)^2} n_k (n_{k'} +1)  \delta(E_b-E_a-\hbar\omega_k+\hbar\omega_{k'})\de \omega_{k} \de \omega_{k'}
\end{equation}
with $C$ given by
\begin{equation}
C = \frac{2\hbar}{\pi} \Big| \langle b |\frac{\partial H_{spin}}{\partial V_{\alpha}} | c\rangle \langle c |\frac{\partial H_{spin}}{\partial V_{\alpha}}| a\rangle \Big|^2.
\end{equation}
Since the contribution are dominant by the peak region, we can make the approximation $\omega_{k}\approx \omega_{\alpha}$ for the denominator. Noting
\begin{equation}
n_{k}(n_{k'}+1)=N(\omega_{k}-\omega_{k'})(n_{k'}-n_{k}),
\end{equation}
with $N(*)$ denoting the Bose-Einstein distribution, we have
\begin{equation}
P_{ba} =  \frac{CN(\omega_{\Delta})}{\hbar(E_c-E_a-\hbar\omega_k)^2}\iint \frac{2\Gamma_{\alpha}}{(\omega_{k'}^2-\omega_{\alpha}^2)^2+\Gamma_{\alpha}^2} \frac{2\Gamma_{\alpha}}{(\omega_k^2-\omega_{\alpha}^2)^2+\Gamma_{\alpha}^2} (n_{k'}-n_{k}) \delta(\omega_{\Delta}-\omega_{k}+\omega_{k'}) \de \omega_{k} \de \omega_{k'},
\end{equation}
where $\hbar\omega_{\Delta}=E_b-E_a$ and $\Gamma_{\alpha}$ is the square of the level broadening of mode $\alpha$. This is similar with the expression for the second order Raman process in Ref.~\cite{Gu2020}, and only the factors out of the integral are different. Following the same argument, we have
\begin{equation}
P_{ba} = \frac{\pi C}{\hbar(E_c-E_a-\hbar\omega_k)^2}\frac{2\omega_{\alpha} \Gamma_{\alpha}}{(\omega_{\alpha}^2\omega_{\Delta})^2+(2\omega_{\alpha}\Gamma_{\alpha})^2}e^{-\omega_{\alpha}/k_BT}.
\end{equation} 
The first order Raman process applies to both non-Kramers and Kramers systems, so is this vibronic barrier formulation. For the latter, $\omega_{\Delta}=0$.

\bibliography{references}
\end{document}